\documentclass[reprint,aps,showpacs,pra,nofootinbib]{revtex4-1}
  \usepackage{color}
  \usepackage{newlfont}
  \usepackage{graphicx}
  \usepackage{amssymb}
  \usepackage{amsmath}
  \usepackage{bm}
  \usepackage{verbatim}
  \usepackage[latin1]{inputenc}
  \usepackage{bbm}
  \usepackage{multirow}
  \usepackage[thinlines]{easytable}
 \usepackage[varg]{txfonts}

\newcommand{\ket}[1]{\mbox{$ | #1 \rangle $}}
\newcommand{\bra}[1]{\mbox{$ \langle #1 | $}}
\newcommand{\be}{\begin{equation}}
\newcommand{\ee}{\end{equation}}
\newcommand{\beq}{\begin{eqnarray}}
\newcommand{\eeq}{\end{eqnarray}}

\begin{document}

\title{Nonanomalous measure of realism-based nonlocality}
\author{V. S. Gomes}
\author{R. M. Angelo}
\affiliation{Department of Physics, Federal University of Paran\'a, P.O. Box 19044, 81531-980 Curitiba, Paran\'a, Brazil}

\begin{abstract}
Based on a recently proposed model of physical reality and an underlying criterion of nonlocality for contexts [A. L. O. Bilobran and R. M. Angelo, Europhys. Lett. {\bf 112}, 40005 (2015)], we introduce a quantifier of realism-based nonlocality for bipartite quantum states, a concept that is profoundly different from Bell nonlocality. We prove that this measure reduces to entanglement for pure states, thus being free of anomalies in arbitrary dimensions, and identify the class of states with null realism-based nonlocality. Then we show that such a notion of nonlocality can be positioned in a low level within the hierarchy of quantumness quantifiers, meaning that it can occur even for separable states. These results open a different perspective for nonlocality studies.

\vskip3mm
\noindent DOI: 10.1103/PhysRevA.97.012123 \hspace{5.6cm} (Published 24 January 2018)
\end{abstract}


\maketitle

\section{Introduction}
Recent experiments~\cite{hensen15,giustina15,shalm15,hensen16,rosenfeld17} have convincingly shown that the correlations measured between far distant systems are not consistent with Bell's hypothesis~\cite{bell64} of local causality\footnote{Although we prefer to employ here the expression ``local causality,'' since it is closer to the one originally used by Bell, ``local realism'' has been the prevalent one in modern approaches. See, e.g., Refs.~\cite{goldstein11,norsen11,gisin12} for critical discussions about this terminology.}
\be \label{lc}
p(a,b|A,B)=\sum_{\lambda}p_{\lambda}\,p(a|A,\lambda)\,p(b|B,\lambda),
\ee 
where $p(a,b|A,B)$ stands for the joint probability of finding outcomes $a$ and $b$ in respective measurements of observables $A$ and $B$, $p(a|A,\lambda)$ and $p(b|B,\lambda)$ are local probability distributions, and $\lambda$ is some hidden variable with probability distribution $p_{\lambda}$. Our current theory of the physical nature, quantum mechanics, correctly predicts the inadequacy of this hypothesis by showing that Bell inequalities grounded on it are violated. Within the context of the quantum-mechanical formalism, the local probability distribution for the part $\cal{A}$ is given by $p(a|A,\lambda)=\text{Tr}(A_a\rho_{\lambda}^{\cal{A}})$, where $\rho_{\lambda}^{\cal{A}}$ is a quantum state in $\cal{H}^{\cal{A}}$ and $A_a$ are projectors defining the spectral decomposition $A=\sum_aaA_a$, with similar expressions for the part $\cal{B}$. Under this restriction, the hypothesis~\eqref{lc} can be written as
\be \label{rhos}
p(a,b|A,B)=\text{Tr}\left(A_a\otimes B_b\,\rho_s\right),\qquad\rho_s=\sum_{\lambda}p_{\lambda}\,\rho_{\lambda}^{\cal{A}}\otimes\rho_{\lambda}^{\cal{B}}.
\ee 
That is, if we strictly confine ourselves to quantum theory, then it is guaranteed that the local causality hypothesis~\eqref{lc} will be readily satisfied by all separable states of the form $\rho_s$. Of course, this is not to say that entanglement and Bell nonlocality are synonymous, for assumption~\eqref{rhos} is only sufficient for local causality. In fact, even though the entanglement-nonlocality equivalence can be claimed for pure states~\cite{gisin91,yu12}, it is well known that there are entangled mixed states that are Bell local~\cite{werner89}. In other words, entanglement is a necessary condition for the observation of Bell nonlocality, but the converse is not true~\cite{brunner14}. Also, there are even subtler instances, referred to as anomalies, where strong Bell inequality violations are achieved with nonmaximally entangled states~\cite{acin02,methot07,vidick11,camalet17}. 

A better understanding of the relation between entanglement and Bell nonlocality is expected to emerge from the analysis of quantifiers of these resources. However, unlike the framework of entanglement quantification, which identifies a set of necessary criteria to be satisfied and counts with many well-established measures for a number of scenarios~\cite{horodecki09}, the research on Bell-nonlocality quantification is still incipient. Approaches known to date involve maximal violations of Bell inequalities~\cite{costa16}, noise resistance~\cite{kas00,las14}, performances in communication tasks~\cite{maudlin92,brassard99,steiner00,bacon03,branciard11}, and, more recently, the volume of violation~\cite{fonseca15}. Remarkably, it has been shown that, via the latter measure, the anomaly claimed for the states $\ket{\psi}=\left(\ket{00}+\gamma\ket{11}+\ket{22}\right)/(2+\gamma^2)^{1/2}$ disappears.

To a certain extent, the magic behind quantum nonlocality can be pictured as follows (see Ref.~\cite{bilobran15} for a thorough argument). Correlations can be viewed as constraints (restrictions) generated by local physical interactions. For example, in preparing the singlet state $\ket{\Psi_s}=\left(\ket{+-}-\ket{-+}\right)/\sqrt{2}$ one establishes the constraint $\cal{S}_z=\cal{S}_z^{\cal{A}}+\cal{S}_z^{\cal{B}}=0$, which implies that the total spin $\cal{S}_z$ is perfectly definite (constrained to zero) whereas the individual spins are not. To see this one can use the basis defined by the total angular momentum and its $z$ component to write $\ket{\Psi_s}=\ket{0,0}$. This result allows us to say that there is an element of reality associated with $\cal{S}_z$, that is, the total spin is real (well defined). In principle, the particles can be sent to different galaxies without degrading the constraint. Now, as soon as particle $\cal{A}$'s spin is measured and thus gets real, the constraint $\cal{S}_z^{\cal{B}}=-\cal{S}_z^{\cal{A}}$  forces particle $\cal{B}$'s spin to become real as well. If the particles are separated by a distance $c\tau+\epsilon$, where $c$ is the speed of light and $\tau$ is the time interval needed for the measurement to be completed, then for all $\epsilon>0$, the establishment of reality elements for the spins is an event that cannot be causally connected, so the notion of nonlocality is inevitable. Within the classical paradigm we can still think of a nonlocally spread constraint, for the angular momentum conservation keeps valid at a distance, but in this case there is no fundamental indefiniteness, that is, the spins are real at every instant of time. This example suggests that irreality (absence of reality elements, indefiniteness) is a basic condition for nonlocality to manifest itself. Now, where exactly does any signature of realism appear in Bell's hypothesis of local causality?

In this work we propose to move the focus away from Bell nonlocality for a while. Instead of considering the local causality hypothesis \eqref{lc} [or \eqref{rhos} in quantum theory], which, arguably, has little to do with realism (see, e.g., Ref.~\cite{gisin12}), we will adopt a given criterion of reality and then ask whether the elements of reality underlying one part of a system can be altered by measurements performed on a distinct remote part of the system. With that, we introduce a quantifier of realism-based nonlocality which is shown to be manifestly nonanomalous and more robust to loss than Bell nonlocality.

\section{Realism-based nonlocality}
We start by indicating the notions of reality and nonlocality that we employ throughout this work. First of all, recall that the Einstein-Podolsky-Rosen (EPR) criterion of reality \cite{EPR35} makes reference to situations where the value of an observable $A=\sum_aaA_a$ can be perfectly predicted without in any way disturbing the system. Clearly, this will be the case if the system state is a projector $A_a=\ket{a}\bra{a}$. For a statistical mixture such as $\rho=\frac{N_1}{N}\ket{1}\bra{1}+\frac{N_3}{N}\ket{3}\bra{3}$, with $N=N_1+N_3$, there is a probability $N_1/N$ of finding the value $a=1$ in a measurement of $A$ and a probability $N_3/N$ of finding $a=3$, so none of the outcomes can be predicted with certainty. A direct application of the EPR criterion to this state would predict ``no element of reality,'' but this would not really be legitimate because this criterion was not originally conceived for dealing with mixed states. In fact, many would agree that $\rho$ is representative of a collection of systems, $N_1$ with an element of reality $a=1$ and $N_3$ with $a=3$. So we could conclude that for this preparation {\em there is} an element of reality for $A$; it is just that we are ignorant about it.

Recently, Bilobran and Angelo (BA) put forward an operational criterion of reality \cite{bilobran15} which agrees with the EPR one in situations involving projectors but also provides a diagnosis in mixed-state contexts. Bilobran and Angelo's criterion is based on the premise that an element of reality is established as soon as a measurement is completed. The protocol is as follows. A given source prepares infinitely many copies of $\rho\in\cal{H}_{\cal{A}}\otimes\cal{H}_{\cal{B}}$. A task is given that consists of determining $\rho$ by ideal state tomography. However, every copy of the system is secretly intercepted by an agent that always measures $A$ and never reveals the outcome. Thus, instead of the preparation $\rho$, the state tomography will reveal the result
\be \label{PhiA}
\Phi_A(\rho):=\sum_aA_a\,\rho\,A_a=\sum_ap_aA_a\otimes\rho_{\cal{B}|a},
\ee 
where $p_a=p(a|A,\rho)=\text{Tr}(A_a\rho)$ and $\rho_{\cal{B}|a}=\text{Tr}_{\cal{A}}(A_a\rho)/p_a$. Because $A$ has been measured and so got real, BA take $\Phi_A(\rho)$ as a state of reality for $A$. This claim is also supported by the fact that a resubmission of the system to the unrevealed-measurement protocol does not change $A$'s reality, that is, $\Phi_A(\Phi_A(\rho))=\Phi_A(\rho)$. Bilobran and Angelo then introduce the  ``irreality $\mathfrak{I}$ of $A$ given $\rho$'' as an entropic distance  to the state of reality $\Phi_A(\rho)$:
\be \label{Ifrak}
\mathfrak{I}(A|\rho):=S(\Phi_A(\rho))-S(\rho),
\ee 
where $S$ is the von Neumann entropy. This is a non-negative quantity that vanishes if and only if $\rho=\Phi_A(\rho)$. Also, we can check that $\mathfrak{I}(A|A_a)=\mathfrak{I}(A|\sum_ap_aA_a)=0$, meaning that the present criterion of reality agrees with the EPR one for eigenstates of $A$ but is more general in the sense that it also predicts full reality for a mixture of $A$'s eigenstates.

With the measure \eqref{Ifrak}, BA propose a notion of nonlocality that is based on alterations of $A$'s reality through physical disturbances occurring in a distant part of the system. Specifically, they consider as an indicator of nonlocality the amount 
\be \label{DI}
\Delta\mathfrak{I}(A,B|\rho)=\mathfrak{I}(A|\rho)-\mathfrak{I}(A|\Phi_B(\rho))
\ee
by which the reality of an observable $A$ accessible in a site $\cal{A}$ changes when unrevealed measurements of an observable $B$ are conducted in a remote site $\cal{B}$. It is shown in Ref.~\cite{bilobran15} that this is a nonnegative quantity which vanishes for uncorrelated states ($\rho=\rho_{\cal{A}}\otimes\rho_{\cal{B}}$) or when an element of reality is already implied by the preparation [$\rho=\Phi_A(\rho)$ or $\rho=\Phi_B(\rho)$]. Also, $\Delta\mathfrak{I}(A,B|\rho)$ is symmetrical under the exchange $A\leftrightarrows B$. It is worth emphasizing that, unlike Bell nonlocality, which derives from violations of the local causality hypothesis~\eqref{lc}, the BA criterion 
\be \label{DI>0}
\Delta\mathfrak{I}(A,B|\rho)>0
\ee 
for the presence of nonlocality in a context defined by $\{A,B,\rho\}$ relies on a notion of realism which is formally characterized by the measure~\eqref{Ifrak}.

\subsection{Quantifying realism-based nonlocality}
Inspired by BA's criterion~\eqref{DI>0} for the nonlocality of a context $\{A,B,\rho\}$, in this work we aim at introducing a measure of the amount of realism-based nonlocality associated solely with a preparation $\rho$. To this end, we take as the realism-based nonlocality quantifier $\cal{N}_{\text{rb}}(\rho)$ the  maximization of the reality variation $\Delta\mathfrak{I}(A,B|\rho)$ over all possible choices of observables $A$ and $B$ acting on $\cal{H_A}$ and $\cal{H_B}$, respectively, that is,
\be \label{Nrb}
\cal{N}_{\text{rb}}(\rho):=\max_{\{A,B\}} \Delta\mathfrak{I}(A,B|\rho).
\ee 
From $\Delta\mathfrak{I}\geqslant 0$ it immediately follows that $\cal{N}_{\text{rb}}(\rho)\geqslant 0$. The motivation behind this definition is twofold. First, it implements the intuition according to which a state that allows for greater changes in the reality of an observable in a site $\cal{A}$ by actions in a remote site $\cal{B}$ is expected to be more nonlocal. Second, $\cal{N}_{\text{rb}}(\rho)$ offers both sufficient and necessary conditions for the presence of realism-based nonlocality, for if $\cal{N}_{\text{rb}}(\rho)=0$, then there is no context $\{A,B,\rho\}$ where the irreality can change, and if $\cal{N}_{\text{rb}}(\rho)>0$, then there is for sure at least one context where the irreality does change. 

{\em No anomaly of realism-based nonlocality.}
In fact, there is a third and more important reason why the above definition is interesting: $\cal{N}_{\text{rb}}$ leads to no anomaly, that is, maximally entangled states $\ket{\Psi_d}=\sum_{i=1}^d\ket{i}\ket{i}/\sqrt{d}$ are diagnosed as maximally nonlocal for all $d$. More specifically, as we prove now, $\cal{N}_{\text{rb}}(\varrho)$ reduces to the entanglement entropy when $\varrho=\ket{\psi}\bra{\psi}$. To start with, we use the definitions~\eqref{Ifrak} and~\eqref{DI} to write the symmetrical form
\be \label{rel1}
\Delta\mathfrak{I}(A,B|\varrho)=S(\Phi_A(\varrho))+S(\Phi_B(\varrho))-S(\Phi_{A,B}(\varrho))-S(\varrho),
\ee 
where $\Phi_{A,B}(\varrho):=\Phi_A(\Phi_B(\varrho))=\Phi_B(\Phi_A(\varrho))$. Since $\varrho$ is pure, then $S(\varrho)=0$. By the Klein inequality~\cite{nielsen00}, one shows that $S(\Phi_A(\varrho))\geqslant S(\varrho)$ and hence that $S(\Phi_{A,B}(\varrho))\geqslant S(\Phi_R(\varrho))$ with $R$ assuming either $A$ or $B$, where the equality holds for $\Phi_{A,B}(\varrho)=\Phi_R(\varrho)$. It follows that 
\be \label{rel2}
2\,S(\Phi_{A,B}(\varrho))\geqslant S(\Phi_A(\varrho))+S(\Phi_B(\varrho)).
\ee 
The relations~\eqref{rel1} and~\eqref{rel2} together imply that
\be 
\Delta\mathfrak{I}(A,B|\varrho)\leqslant \tfrac{1}{2}\left[S(\Phi_A(\varrho))+S(\Phi_B(\varrho))\right],
\ee 
with the equality holding for $\Phi_{A,B}(\varrho)=\Phi_A(\varrho)=\Phi_B(\varrho)$. This inequality can be saturated for observables $A$ and $B$ that define the Schmidt basis of $\ket{\psi}$. To see this, take the Schmidt decomposition $\ket{\psi}=\sum_i\sqrt{\xi_i}\ket{\alpha_i}\ket{\beta_i}$ and the Schmidt observables $\alpha=\sum_i\alpha_i\ket{\alpha_i}\bra{\alpha_i}$ and $\beta=\sum_i\beta_i\ket{\beta_i}\bra{\beta_i}$. By direct calculation one shows that $\Phi_{\alpha}(\varrho)=\Phi_{\beta}(\varrho)=\sum_i\xi_i\ket{\alpha_i}\bra{\alpha_i}\otimes\ket{\beta_i}\bra{\beta_i}$, which implies that $\Phi_{\alpha,\beta}(\varrho)=\Phi_{\alpha}(\varrho)=\Phi_{\beta}(\varrho)$. Hence, the maximization in Eq.~\eqref{Nrb} is attained with $A=\alpha$ and $B=\beta$, that is, 
\be \label{Nalphabeta}
\cal{N}_{\text{rb}}(\varrho)=\Delta\mathfrak{I}(\alpha,\beta|\varrho)=\tfrac{1}{2}\left[S(\Phi_{\alpha}(\varrho))+S(\Phi_{\beta}(\varrho))\right].
\ee 
Let $E(\varrho):=S(\text{Tr}_{\cal{A}}(\varrho))=S(\text{Tr}_{\cal{B}}(\varrho))$ be the entanglement entropy of $\varrho=\ket{\psi}\bra{\psi}$. Here it results in $E(\varrho)=-\sum_i\xi_i\ln\xi_i$. By use of the joint-entropy theorem \cite{nielsen00}, one may show that $S(\Phi_{\alpha}(\varrho))=S(\Phi_{\beta}(\varrho))=E(\varrho)$. By plugging this result into Eq.~\eqref{Nalphabeta} we finally prove the point:
\be 
\cal{N}_{\text{rb}}(\varrho)=E(\varrho) \qquad\qquad\left(\varrho=\ket{\psi}\bra{\psi}\right).
\ee 
In other words, for bipartite pure states the realism-based nonlocality equals entanglement. In particular, for the maximally entangled state $\ket{\Psi_d}$ our quantifier reaches it maximum, namely, $\cal{N}_{\text{rb}}(\ket{\Psi_d})=\ln d$. Remarkably, this shows that no anomaly will emerge for this measure in any dimension $d$.

\subsection{Hierarchy of quantumness measures}
As discussed in Ref.~\cite{bilobran15}, $\Delta\mathfrak{I}=0$ for both fully uncorrelated states $(\rho=\rho_{\cal{A}}\otimes\rho_{\cal{B}})$ and states of reality [$\rho=\Phi_A(\rho)$ or $\rho=\Phi_B(\rho)$]. By~Eq. \eqref{PhiA} we see that states of reality are necessarily separable. Thus, the most general state with element of reality for $A$ can be written as
\be 
\rho_{[A]}=\sum_kp_k\,\Phi_A(\rho_k^{\cal{A}})\otimes\rho_k^{\cal{B}},
\ee 
which clearly satisfies $\Phi_A(\rho_{[A]})=\rho_{[A]}$. This reveals that such states require, beyond separability, full reality for its reduced form $\text{Tr}_{\cal{B}}\rho_{[A]}$. On the other hand, they are less restrictive than states like $\sum_kp_k A_k\otimes\rho_k^{\cal{B}}$ and $\sum_kp_kA_k\otimes B_k$, for which the one-way quantum discord~\cite{ollivier01,henderson01} and the symmetric quantum discord~\cite{rulli11} vanish, respectively. This observation allows us to position the realism-based nonlocality~\eqref{Nrb} within a hierarchy of quantumness measures \cite{wiseman07,costa16}. This is done in Table~\ref{table1}, where many quantumness measures are organized in terms of the models that negate the existence of those classes of quantumness. The construction of such hierarchy is supported by the following rationale.

To check whether a context displays Bell nonlocality we take the task of describing the experimental joint probability distribution $p(a,b|A,B)$ in terms of the local causal model~\eqref{lc} with no restriction at all for the local probability distributions $p(a|A,\lambda)$ and $p(b|B,\lambda)$, where $\lambda$ plays the role of a generic hidden variable (an arbitrary physical state or parameter) that determines those distributions. Finding these local distributions we prove the absence of Bell nonlocality. Now, to prove the absence of one-way EPR steering, besides taking the local causal model~\eqref{lc} we impose a restriction, namely, that the local distribution $p(a|A,\lambda)$ be strictly constructed in consistency with quantum mechanics. In this case, we have $p(a|A,\lambda)=\text{Tr}(A_a\rho_{\lambda}^{\cal{A}})$, so that the generic variable $\lambda$ gets restricted to the class of quantum states $\rho_{\lambda}^{\cal{A}} \in \cal{H_A}$. To prove the absence of entanglement, we add another restriction: confining $p(b|B,\lambda)$ as well to the quantum formalism. With that, we get restricted to models like~\eqref{rhos} which, if proved to exist, negate entanglement. Realism-based nonlocality is probed by further restricting the reduced states of part $\cal{A}$ to states of reality $\Phi_{A'}(\rho_{\lambda}^{\cal{A}})$, for any $A'$ acting on $\cal{H_A}$. In this case, by setting $A'=A$ one attempts to describe the experimental distribution as $p(a,b|A,B)=\text{Tr}(A_a\otimes B_b\,\rho_{[A]})$. If this is shown to be possible, then no realism-based nonlocality will be found. Finally, by restricting the reality states to the projectors $A_{\lambda}'$ and $B_{\lambda}'$ we can probe the variants of quantum discord.

Since many models can be found to satisfy the Bell locality hypotheses given in Table~I (local causality plus any local probability distributions) then the set of Bell nonlocal states is expected to be relatively small, that is, Bell nonlocality is the most restrictive class of quantumness. At the other end, the symmetric quantum discord appears: Since the hypotheses for its absence are very specific (local causality plus local distributions consistent with projectors), the number of symmetrically discordant states is expected to be huge. It then follows that Bell nonlocal states form a strict subset of EPR steerable states, which forms a strict subset of entangled states, which forms a strict subset of realism-based nonlocal states, which forms a strict subset of discordant states, which in turn forms a strict subset of symmetrically discordant states. Alternatively, one may quote this quantumness hierarchy by saying that there is an ordering of implication as we move upward along the third column of Table~\ref{table1}. 

\begin{table}[htb]
\caption{Hierarchy of quantumness measures. The first two columns define the structures presumed for the local probability distributions which, along with the Bell hypothesis of local causality, $p(a,b|A,B)=\sum_{\lambda}p_{\lambda}p(a|A,\lambda)p(b|B,\lambda)$, imply the absence of the corresponding quantumness class given in the third column. In the first two rows, the symbol $\forall$ indicates that any state or parameter $\lambda$ of any hidden variable theory is admitted. For the other cases, only quantum states $\rho_{\lambda}^{\cal{R}}\in\cal{H_{R}}$ (with $\cal{R}$ assuming either $\cal{A}$ or $\cal{B}$) are considered. Here $A$ and $A'$, and their corresponding projectors $A_a$ and $A'_{\lambda}$, refer to arbitrary observables acting on $\cal{H_A}$, with similar interpretations for $B_b$ and $B'_{\lambda}$.}
\setlength{\tabcolsep}{5pt} 
\begin{tabular}{lcc}\hline\hline
$p(a|A,\lambda)$ & $p(b|B,\lambda)$ &  Absent quantumness  \\ \hline 
$\forall$ & $\forall$ & Bell nonlocality  \\ 
$\text{Tr}\left(A_a\rho_{\lambda}^{\cal{A}}\right)$  & $\forall$ & one-way EPR steering  \\ 
$\text{Tr}\left(A_a\rho_{\lambda}^{\cal{A}}\right)$ & $\text{Tr}\left(B_b\rho_{\lambda}^{\cal{B}}\right)$ & entanglement  \\ 
$\text{Tr}\left[A_a\Phi_{A'}(\rho_{\lambda}^{\cal{A}})\right]$ & $\text{Tr}\left(B_b\rho_{\lambda}^{\cal{B}}\right)$ & realism-based nonlocality  \\ 
$\text{Tr}\left(A_a A'_{\lambda} \right)$ & $\text{Tr}\left(B_b\rho_{\lambda}^{\cal{B}}\right)$ & one-way quantum discord \\ 
$\text{Tr}\left(A_a A'_{\lambda} \right)$ & $\text{Tr}\left(B_bB'_{\lambda}\right)$ & symmetric quantum discord \\ \hline\hline
\end{tabular} 
\label{table1}
\end{table} 

\subsection{Comparison with other nonlocality measures}
Here we analytically compute the realism-based nonlocality $\cal{N}_{\text{rb}}$, via Eqs.~\eqref{DI} and~\eqref{Nrb}, for the family of two-qubit Werner states
\be\label{rhomu}
\rho_{\mu}=(1-\mu)\tfrac{\mathbbm{1}}{4}+\mu\,\ket{\Psi_s}\bra{\Psi_s},
\ee 
where $\ket{\Psi_s}$ is the singlet state and $\mu\in[0,1]$. With the $\rho_{\mu}$ eigenvalues  $\{\tfrac{1-\mu}{4},\tfrac{1-\mu}{4},\tfrac{1-\mu}{4},\tfrac{1+3\mu}{4}\}$, we compute $S(\rho_{\mu})$. The maps $\Phi_{A,B}$ are constructed for the observables $A=\hat{u}\cdot\vec{\sigma}$ and $B=\hat{v}\cdot\vec{\sigma}$ (whose eigenvalues and projectors can be easily determined), with the unit vectors $\{\hat{u},\hat{v}\}$ and the Pauli matrices $\vec{\sigma}=(\sigma_1,\sigma_2,\sigma_3)$. We then obtain the eigenvalues $\tfrac{1\pm\mu}{4}$ (each one being twofold degenerate) for both the reality states $\Phi_A(\rho_{\mu})$ and $\Phi_B(\rho_{\mu})$. The eigenvalues of $\Phi_{A,B}(\rho_{\mu})$ can be shown to be $\tfrac{1}{4}(1\pm\mu |\hat{u}\cdot\hat{v}|)$, each one with degeneracy 2. Then, by writing $|\hat{u}\cdot\hat{v}|:=\eta\in[0,1]$ we reduce the maximization process to the computation of $\min_{\eta}S(\Phi_{A,B}(\rho))$, whose result emerges for $\eta=1$. Putting all this together we find
\be \label{Nrb_res}
\cal{N}_{\text{rb}}(\rho_{\mu})=\tfrac{1}{4}\Big[h(3\mu)+h(-\mu)-2h(\mu)\Big],
\ee
where $h(\mu):=(1+\mu)\ln{(1+\mu)}$. In what follows, for the sake of comparison, we compute other two quantifiers of Bell nonlocality. We start with the recently introduced volume of violation~\cite{fonseca15}. Consider the Clauser-Horne-Shimony-Holt (CHSH) inequality 
\be \label{CHSH}
\mathbb{B}:=\left| A_1B_1+A_1B_2+A_2B_1-A_2B_2\right|\leqslant 2,
\ee 
where $A_iB_j\equiv\text{Tr}\left[\rho\left(\hat{u}_i\cdot\vec{\sigma}\otimes\hat{v}_j\cdot\vec{\sigma}\right)\right]$ with real unit vectors $\hat{u}_{1,2}$ and $\hat{v}_{1,2}$. For the state \eqref{rhomu} one finds that $A_iB_j=-\mu\,\hat{u}_i\cdot\hat{v}_j$ and $\mathbb{B}=\mu\,\left|\hat{u}_1\cdot\left(\hat{v}_1+\hat{v}_2 \right)+\hat{u}_2\cdot\left(\hat{v}_1-\hat{v}_2 \right)\right|$. Thus, the condition of violation of the inequality~\eqref{CHSH} becomes
\be \label{violation}
\left|\hat{u}_1\cdot\left(\hat{v}_1+\hat{v}_2\right)+\hat{u}_2\cdot\left(\hat{v}_1-\hat{v}_2 \right)\right|>2/\mu.
\ee 
If the unit vectors are parametrized with spherical angles, the computation of the volume of violation demands counting the number of points $\{\theta_{u_1},\phi_{u_1},\theta_{u_2},\phi_{u_2},\theta_{v_1},\phi_{v_1},\theta_{v_2},\phi_{v_2}\}$ that satisfy the condition~\eqref{violation}. This constitutes a rather complicated problem, with analytical solution known only for a very specific scenario~\cite{parisio16} (corresponding here to $\mu=1$). To circumvent this difficulty, we introduce the real variables 
\be \label{xyz}
x=\frac{\hat{u}_1\cdot(\hat{v}_1+\hat{v}_2)}{||\hat{v}_1+\hat{v}_2||},\,\, y=\frac{\hat{u}_2\cdot(\hat{v}_1-\hat{v}_2)}{||\hat{v}_1-\hat{v}_2||},\,\, z=\frac{1+\hat{v}_1\cdot\hat{v}_2}{2},
\ee 
such that $\{x,y\}\in[-1,1]$ and $z\in[0,1]$. This strategy greatly reduces the dimensionality of the integration space; for now the violation condition~\eqref{violation} can be simply expressed as
\be \label{muB>1}
\mu\,\mathfrak{B}(x,y,z)> 1,\qquad \mathfrak{B}(x,y,z):=\left|x\sqrt{z}+y\sqrt{1-z}\right|.
\ee 
The volume of violation for the state~\eqref{rhomu} then reads
\be \label{Nvol}
\cal{N}_{\text{vol}}(\rho_{\mu})=\frac{1}{V}\int\int\int_{\Gamma}\,\mathrm{d}x\,\mathrm{d}y\,\mathrm{d}z,
\ee 
where $\Gamma$ is the subspace of points $\{x,y,z\}$ that satisfies the condition~\eqref{muB>1} and $V=4$ is the total volume. Before proceeding with the calculation of these integrals, we prove that $\cal{N}_{\text{vol}}(\rho_{\mu})=0$ for $\mu\leqslant 1/\sqrt{2}$. With the parametrization $\mu=(1+\epsilon)/\sqrt{2}$ we express the violation condition as $\mathfrak{B}>\sqrt{2}/(1+\epsilon)$.  From $(X-Y)^2\geqslant 0$ we get $2XY\leqslant X^2+Y^2$, which allows us to show that $|X+Y|\leqslant [2(X^2+Y^2)]^{1/2}$. Setting $Y=y\sqrt{1-z}$ and $X=x\sqrt{z}$, one obtains $\mathfrak{B}=|X+Y|\leqslant \sqrt{2}$. It then follows that $\sqrt{2}/(1+\epsilon)<\sqrt{2}$, which is satisfied only if $\epsilon>0$, that is, for $\mu>1/\sqrt{2}$.

\begin{figure}[htb]
\includegraphics[width=\columnwidth]{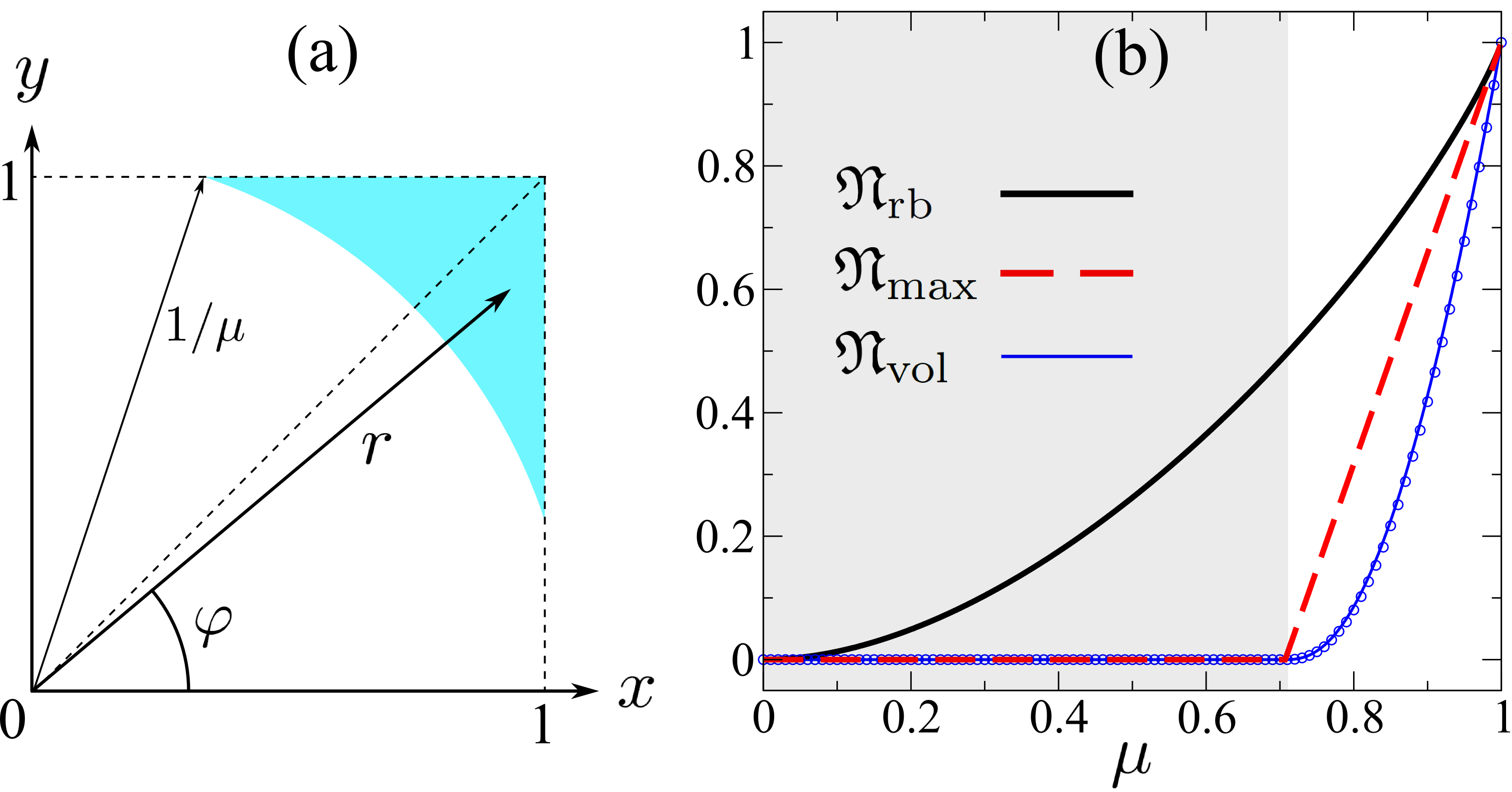}
\caption{(a) One quarter of the $xy$ domain (cyan shaded area) where the violation condition \eqref{muB>1} [or \eqref{mur>1}] is satisfied. A detailed inspection of this geometry allows for the analytical computation of the volume of violation \eqref{Nvol}. (b) Normalized nonlocality quantifiers $\mathfrak{N}_{\text{rb}}$ (thick black line), $\mathfrak{N}_{\max}$ (dashed red line), and $\mathfrak{N}_{\text{vol}}$ (thin blue line) for the state \eqref{rhomu} as a function of the parameter $\mu\in[0,1]$. The small blue circles refer to numerical results obtained for $\mathfrak{N}_{\text{vol}}$ by counting points $\{\theta_{u_1},\phi_{u_1},\theta_{u_2},\phi_{u_2},\theta_{v_1},\phi_{v_1},\theta_{v_2},\phi_{v_2}\}$ that satisfy the violation condition \eqref{violation}. In the gray region $(\mu\leqslant 1/\sqrt{2})$ the CHSH inequality predicts no Bell nonlocality. For $\mu>1/\sqrt{2}$, the Bell-nonlocality quantifiers are monotonically increasing functions of $\mu$ and, in this sense, are equivalent.}
\label{fig1}
\end{figure}

It is easy to check that $\mathfrak{B}$ is maximized if $z=x^2/(x^2+y^2)$. This leads to $\mathfrak{B}\leqslant (x^2+y^2)^{1/2}$, which implies that $x^2+y^2>\mu^{-2}$. Introducing the parametrization
\be \label{phi_theta}
x=r\cos{\varphi},\qquad y=r\sin{\varphi}, \qquad z=\sin^2{\Theta},
\ee 
with $\varphi\in [0,2\pi]$, $\Theta\in[0,\tfrac{\pi}{2}]$, and $\mathrm{d}x\mathrm{d}y\mathrm{d}z=r\sin(2\Theta)\mathrm{d}r\mathrm{d}\varphi\mathrm{d}\Theta$, the latter inequality and the condition~\eqref{muB>1} are respectively written as
\be \label{mur>1}
r>1/\mu, \qquad\qquad r\,\left|\sin(\Theta+\varphi)\right|>1/\mu.
\ee 
From the second inequality we derive the integration limits for $\Theta$ as a function of $r$ and $\varphi$. The first inequality allows one to infer the integration domain in the $xy$ plane, as is depicted in Fig.~\ref{fig1}(a). By symmetry, we can focus on the region below the line $\varphi=\tfrac{\pi}{4}$, where the integration limits are $\{\mu^{-1},(\cos{\varphi})^{-1}\}$ for $r$ and $\{\arccos{\mu},\tfrac{\pi}{4}\}$ for $\varphi$. With all this, we can perform the integrations in Eq.~\eqref{Nvol} to obtain
\be \label{Nvol_res}
\cal{N}_{\text{vol}}(\rho_{\mu})=\frac{8}{3}\left(\mu+\frac{1}{\mu}\right)g_{\mu}+\frac{2}{\mu}\ln\left[4\mu^2\left(1-g_{\mu} \right)-1\right],
\ee
where $g_{\mu}:=[1-1/(2\mu^2)]^{1/2}$. We have checked the validity of this result by numerically counting, within a random sampling of $10^{8}$ points $\{\theta_{u_1},\phi_{u_1},\theta_{u_2},\phi_{u_2},\theta_{v_1},\phi_{v_1},\theta_{v_2},\phi_{v_2}\}$, the ones satisfying the condition \eqref{violation}. No significant deviations were observed within the margin of error of the statistical method adopted [see Fig.~\ref{fig1}(b)].

Finally, we compute $\cal{N}_{\max}(\rho_{\mu})$, which consists of the amount by which the CHSH inequality is maximally violated for the state $\rho_{\mu}$. Mathematically, this is written as
\be 
\cal{N}_{\max}(\rho_{\mu})=\max\left[0,\max_{\{x,y,z\}} \mu\mathfrak{B}(x,y,z)-1\right].
\ee 
We have seen that $z=x^2/(x^2+y^2)$ saturates $\mathfrak{B}$ to $(x^2+y^2)^{1/2}$, which then reaches its maximum for $x=y=1$. Hence,
\be \label{Nmax_res}
\cal{N}_{\max}(\rho_{\mu})=\max\left[0,\mu\sqrt{2}-1\right],
\ee 
which implies no Bell nonlocality for $\mu\leqslant 1/\sqrt{2}$, in total agreement with both the $\cal{N}_{\text{vol}}(\rho_{\mu})$ predictions and (up to a normalization factor) the result reported in Ref.~\cite{costa16PRA}.

The results \eqref{Nrb_res}, \eqref{Nvol_res}, and \eqref{Nmax_res} are monotonically increasing functions of $\mu$ in the domain $(\tfrac{1}{\sqrt{2}},1]$. They all reach their maximum values for the singlet state $\rho_{\mu=1}$ and thus agree in that the maximally entangled two-qubit state is maximally nonlocal. Figure \ref{fig1}(b) provides a comparison among the quantifiers in terms of their respective normalized forms, which are defined as
\be 
\mathfrak{N}_{\square}(\rho_{\mu})=\frac{\cal{N}_{\square}(\rho_{\mu})}{\cal{N}_{\square}(\rho_{\mu=1})}\qquad (\square=\text{rb, vol, max}).
\ee 
What is most remarkable about these results is that the realism-based nonlocality extends over the whole domain $[0,1]$ of $\mu$, being present even when no Bell nonlocality occurs. Along with the hierarchy predicted in Table~\ref{table1}, this suggests that $\cal{N}_{\text{rb}}$ must occur for almost all quantum states. Also, if we set $\mu=e^{-t}$, for some dimensionless time scale $t$, we may think of $\rho_{\mu}$ as modeling a singlet state subjected to a noisy dynamics. In this case, we see that while $\mathfrak{N}_{\max}$ and $\mathfrak{N}_{\text{vol}}$ eventually undergo sudden death, $\mathfrak{N}_{\text{rb}}$ vanishes only asymptotically, thus being notably more robust to lossy channels.

\section{Concluding remarks} 
In this work, we have taken a different premise for nonlocality. Instead of employing Bell's hypothesis of local causality, we adopted a notion of reality which has recently been formulated on both quantitative and operational grounds~\cite{bilobran15}. This strategy allowed us to introduce the realism-based nonlocality~ \eqref{Nrb}, which is associated with the amount by which the reality in a given site maximally changes via measurements conducted in a remote site. Besides reducing to the entanglement entropy for bipartite pure states of arbitrary dimensions (thus being anomaly-free), our measure occupies a position within the hierarchy of quantumness quantifiers that makes it ubiquitous over the space of quantum preparations. An interesting perspective to be exploited is whether the realism-based nonlocality can play some relevant role in the contexts of resource theories, communication tasks, and foundational debates.

\section*{Acknowledgments} 
V.S.G. and R.M.A. respectively acknowledge financial support from CAPES (Brazil) and the National Institute for Science and Technology of Quantum Information (CNPq, Brazil).


\end{document}